\documentclass[journal=jacsat,manuscript=article]{achemso}

\usepackage[version=3]{mhchem} 
\usepackage{physics}
\usepackage{parskip} 
\usepackage[utf8]{inputenc}
\usepackage{subcaption}
\usepackage[toc,page]{appendix}
 \SectionNumbersOn

\author{Rishabh Gupta}
\author{Rongxin Xia}
\affiliation[Purdue University]
{Department of Chemistry, Department of Physics and Astronomy,  Purdue University, West Lafayette, IN, USA}
\author{Raphael D. Levine}
\email{raphy@mail.huji.ac.il}
\affiliation[Hebrew University]
{The Fritz Haber Center for Molecular Dynamics and Institute of Chemistry, The Hebrew University of Jerusalem, Jerusalem 91904, Israel
}
\alsoaffiliation[University of California]
{Department of Chemistry and Biochemistry and  Department of Molecular and
Medical Pharmacology, David Geffen School of Medicine, University of California, Los
Angeles, CA 90095, USA}

\author{Sabre Kais}
\email{kais@purdue.edu}
\affiliation[Purdue University]
{Department of Chemistry, Department of Physics and Astronomy, and Purdue Quantum Science and Engineering Institute, Purdue University, West Lafayette, IN, USA}

\title[An \textsf{achemso} demo]
  {Maximal entropy approach for quantum state tomography}

\setcitestyle{numbers}
\mciteErrorOnUnknownfalse
\DeclareUnicodeCharacter{2212}{-}
\begin{document}








\begin{abstract}
Quantum computation has been growing rapidly in both theory and experiments. In particular, quantum computing devices with a large number of qubits have been developed by IBM, Google, IonQ, and others. The current quantum computing devices are noisy intermediate-scale quantum (NISQ) devices, and so approaches to validate quantum processing on these quantum devices are needed. One of the most common ways of validation for an \textit{n}-qubit quantum system is quantum tomography, which tries to reconstruct a quantum system’s density matrix by a complete set of observables. However, the inherent noise in the quantum systems and the intrinsic limitations poses a critical challenge to precisely know the actual measurement operators which make quantum tomography  impractical in experiments. Here, we propose an alternative approach to quantum tomography, based on the maximal information entropy, that can predict the values of unknown observables based on the available mean measurement data. This can then be used to reconstruct the density matrix with high fidelity even though the results for some observables are missing. Of additional contexts, a practical approach to the inference of the quantum mechanical state using only partial information is also needed.
\end{abstract}

\section{Introduction}
Quantum technologies increasingly require measuring the mean value of observables towards determining a detailed state of a nano-system. The method of characterization of a quantum system [\cite{nielson},\cite{kais},\cite{photonic},\cite{qubits},\cite{cramer}] by measuring the expectation values of a complete set of observables is known as quantum state tomography [\cite{photonic},\cite{cramer},\cite{quasi}] and forms an important basis in quantum computation operations [\cite{nielson},\cite{evange},\cite{maxentropy}]. A simple example is the case when the system can be described by a wave function spanning \textit{N} basis states: $\ket{\psi} = \sum_{i=1}^{\textit{N}} a_{i}\ket{i}$. The \textit{a$_{i}$}'s are complex valued amplitudes and $\abs{\textit{a}_i}^2$ = $\abs{\braket{i}{\psi}}^2$ is the probability to be in the quantum basis $\ket{\textit{i}}$. The exact density matrix can be constructed using the \textit{$N(N-1)/2$} independent complex numbers S$_{ij}$ = $a_{i}a_{j}^{*}$ that are equivalent to \textit{$N^2$-1} real numbers and these real numbers are observables as they are the expectation values of Hermitian operators, $a_{i}a_{j}^{*}$+($a_{i}^{*}a_{j}$)$^{*}$ = $\bra{\psi}(\ket{i}\bra{j}+\ket{j}\bra{i})\ket{\psi}$. The finite values of the observables generated by the dynamics allow us to specify \textit{N$^{2}$-1} independent unitary matrices which gives rise to SU(N) Lie algebra. SU(2), for example, is a set of three unitary matrices and is what is generated by a qubit. The SU(2) algebra that describes a two state quantum mechanical system is the quantum generalization of Boolean algebra and can be realized by its three Pauli generators ($\sigma_x$, $\sigma_y$ and $\sigma_z$). \par
One would expect on general grounds [\cite{james2005measurement}] that if one could make repeated measurements it will be possible to characterize the quantum mechanical state with fewer than \textit{N$^2$} observables. Like the classical noiseless coding theorem of Shannon [\cite{nielson},\cite{shannon}] the quantum mechanical version by Schumacher [\cite{nielson},\cite{schumacher}] does not tell us how to construct a code that will do it. This is one motivation for trying to use fewer than \textit{N$^2$} observables. The other is that in a variety of practical contexts we have a limited number of observations or we have measurements of limited fidelity. This is particularly so when the quantum device is noisy as in the so called intermediate-scale quantum (NISQ) devices [\cite{preskill2018quantum}]. This may lead to an ambiguous description of a quantum state. \par
Recently, many new approaches have been proposed to implement quantum tomography, by assuming the quantum system is in a low-rank state to reduce unnecessary measurements [\cite{gross2010quantum}] or use the current measurement results to decide the choice of next measurements [\cite{ferrie2014self}]. Here we propose to use a maximization of the von Neumann entropy [\cite{buzek1998quantum}] searching for a maximum of the entropy of the state constrained by the given value of the observations at hand. See [\cite{wich},\cite{jaynes},\cite{katz1967principles}] for a general discussion of this point of view. Here we use a special kind of constraints, the measured mean values of populations and coherences. In other words, our constraints will be an incomplete set of the generators of a unitary algebra when normalization is imposed. \par
In addition to its inherent interest, quantum state tomography has applications in a variety of different fields such as characterization of optical signals [\cite{ariano}], validation of quantum gates in varied quantum computing operations [\cite{nielson},\cite{kais},\cite{photonic}], studying the dynamics of quantum states via quantum process tomography [\cite{aspuru},\cite{aspuru2},\cite{aspuru3}] and applications to computing by observables [\cite{fresch},\cite{Gattuso2020},\cite{fresch2}]. \par
In this paper we discuss an alternative approach to reconstruct the density matrix based on the maximal information entropy [\cite{wich},\cite{jaynes},\cite{katz1967principles},\cite{raphy}] and use a finite but incomplete set of observables. We demonstrate an explicit scheme for generating a density matrix for the common quantum mechanical situation when the different observables do not commute. To test the reconstructed density matrix, we carry out the state tomography by using measurements from quantum circuits implemented on IBM quantum computing chips. These quantum chips are composed of superconducting transmon qubits and can be easily accessed through IBM quantum experience [\cite{SANTOS2017}]. We apply our proposed reconstruction to the cases where the observables are a few members of the generators of the SU(4) and SU(8) algebras. We intend to further study this approach for more complicated cases, for example, SU(N) for higher values of \textit{N}, and compare the density matrix reconstruction based on different number of known mean values of observables.

\section{Maximal information entropy-based density reconstruction}
Concatenating two qubit units leads to an SU(4) algebra. We start with this simple example of a 2-qubit quantum system. In this case, where 4 basis states are involved in the dynamics, an exact complete description requires measurements of the expectation values of 16 operators:
\begin{eqnarray}
    &\hspace*{0.001cm}&\{\ket{1}\bra{1},\ket{2}\bra{2},\ket{3}\bra{3},\ket{4}\bra{4},(\ket{1}\bra{2},\ket{2}\bra{1}),(\ket{1}\bra{3},\ket{3}\bra{1}),(\ket{1}\bra{4},\ket{4}\bra{1}), \nonumber \\
    &\hspace*{0.001cm}&(\ket{2}\bra{3},\ket{3}\bra{2}), (\ket{2}\bra{4},\ket{4}\bra{2}), (\ket{3}\bra{4},\ket{4}\bra{3}) \} \label{basis1}    
\end{eqnarray}
The first four of those are probabilities of the four basis states and the twelve others correspond to the coherences between them. Following the formalism of the maximal information entropy [\cite{raphy}] subject to known average values of certain operators $\hat{f}_{k}$, the density operator in terms of the Lagrange multipliers $\lambda_k$ [\cite{raphy2}] can be obtained as [\cite{wich},\cite{jaynes},\cite{Green1450}]:
\begin{eqnarray}
\hat{\rho} = \frac{1}{Z(\lambda_{1},\ldots,\lambda_{k})}\exp\{-\sum_{k}\lambda_{k}\hat{f}_{k}\} \label{rho2}
\end{eqnarray}
where $Z(\lambda_{1},\ldots,\lambda_{k})=Tr(\exp\{-\sum_{k}\lambda_{k}\hat{f}_{k}\})$ insures the normalization as $Tr(\rho)=1$. Eq. (\ref{rho2}) already assumes that not all observables are available, so let us consider the case where we only know the mean values of two probabilities and a coherence. The density operator is then defined as:
\begin{eqnarray}
\hat{\rho} = \frac{1}{Z(\lambda_{11},\lambda_{12},\lambda_{22})}\exp\{-\lambda_{11}\ket{1}\bra{1}-\lambda_{12}\ket{1}\bra{2} -\lambda_{12}^{*}\ket{2}\bra{1}-\lambda_{22}\ket{2}\bra{2}\} \label{abc00}
\end{eqnarray}
To compute the density matrix explicitly we first diagonalize the Hermitian matrix \textbf{A} which is the exponent in Eq. (\ref{abc00}):
\[
A = 
\begin{bmatrix}
 -\lambda_{11} & -\lambda_{12} & 0 & 0  \\
  -\lambda_{12}^{*} & -\lambda_{22} & 0 & 0 \\
  0 & 0 & 0 & 0 \\
  0 & 0 & 0 & 0 
 \end{bmatrix} =\sum_{i=1}^{4} \epsilon_{i}\ket{\phi_{i}}\bra{\phi_{i}}
\]
$\{\epsilon_i,\ket{\phi_i}\}$ correspond to the eigenvalues and eigenvectors of A. The density operator can now be computed in the basis of eigenvectors of A:
\begin{eqnarray}
\hat{\rho} &=& \frac{1}{Z}\left( \exp{\epsilon_{1}}\ket{\phi_{1}}\bra{\phi_{1}}+\exp{\epsilon_{2}}\ket{\phi_{2}}\bra{\phi_{2}}+\exp{\epsilon_{3}}\ket{\phi_{3}}\bra{\phi_{3}}+\exp{\epsilon_{4}}\ket{\phi_{4}}\bra{\phi_{4}}\right) \label{rhoh} \\
Z &=& tr(\exp{\textbf{A}}) = \sum_{i=1}^{4} \exp{\epsilon_{i}}
\end{eqnarray}
A practical form of Eq. (\ref{rhoh}) is obtained by expanding the projection operators $\ket{\phi_{i}}\bra{\phi_{i}}$ in terms of our initial basis (\ref{basis1}) and thereby giving the final form of the density operator:
\begin{eqnarray}
\hat{\rho}&=&\frac{1}{Z}\sum_{i}\exp{\epsilon_{i}}|\phi_{i}><\phi_{i}| \nonumber \\
&=&\frac{1}{Z}(|4><4|+|3><3|+(a+b)|1><1|+(\frac{a}{k_{3}^*}+\frac{b}{k_{4}^*})|1><2| \nonumber\\
&+&(\frac{a}{k_{3}}+\frac{b}{k_{4}})|2><1|+(\frac{a}{\abs{k_{3}}^{2}}+\frac{b}{\abs{k_{4}}^{2}})|2><2|) \label{density1}
\end{eqnarray}
where Z=$\sum_{i}\exp{\epsilon_{i}}$, k$_3$ = -($\frac{\epsilon_3}{\lambda_{12}^{*}}$+$\frac{\lambda_{22}}{\lambda_{12}^{*}}$), k$_4$ = -($\frac{\epsilon_4}{\lambda_{12}^{*}}$+$\frac{\lambda_{22}}{\lambda_{12}^{*}}$), a=$\frac{\abs{k_{3}}^{2}}{\sqrt{(k_3^2+1)({k_3^*}^2+1)}}\exp{\epsilon_{3}}$, and \par b=$\frac{\abs{k_{4}}^{2}}{\sqrt{(k_4^2+1)({k_4^*}^2+1)}}\exp{\epsilon_{4}}$. \par
\begin{figure}[ht!]
  \centering 
  \includegraphics[width=3.2in]{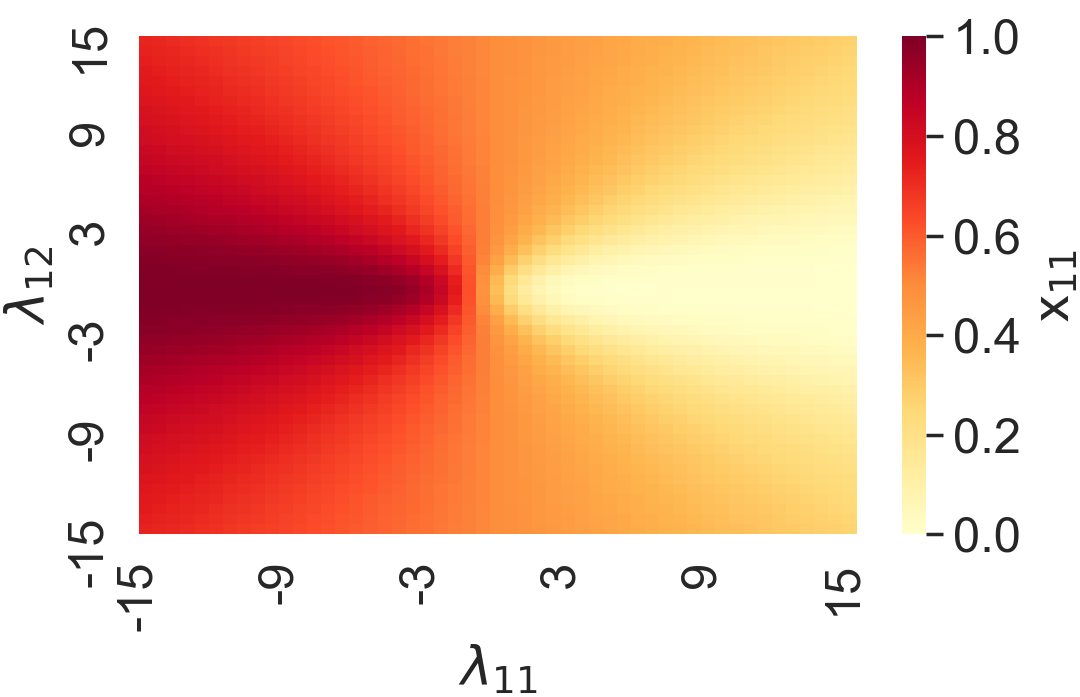} \hspace*{\fill}
    \includegraphics[width=3.2in]{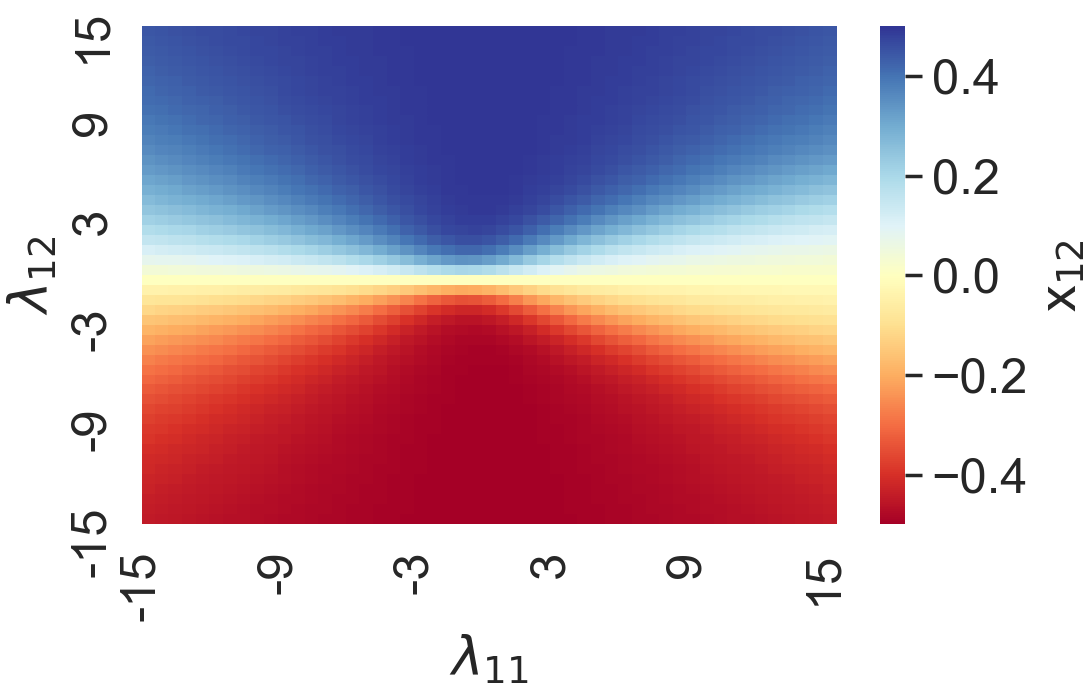}
\caption{Heat maps representation of $x_{11}$ and $x_{12}$ as a function of $\lambda_{11}$ and $\lambda_{12}$ (keeping Im($\lambda_{12}$) = 0, $\lambda_{22}$ = 0). If there are no constraints then all three quantum states should be equally probable so that x$_{11}$ = 1/3 and there should be no finite coherence so that x$_{12}$ = 0. This is seen to be the case when the Lagrange multipliers are all zero. From these maps one can read the values of $\lambda_{11}$ and $\lambda_{12}$ for each given value of the observables.}
\label{fig_1}
\end{figure}

 
To determine the values of the Lagrange multipliers $\lambda_{ij}$ we use the information about the measured mean values of the operators $x_{11}$= $\langle\ket{1}\bra{1}\rangle$, $x_{12}$= $\langle\ket{1}\bra{2}\rangle$, and $x_{22}$= $\langle\ket{2}\bra{2}\rangle$. Based on the density operator defined by Eq. (\ref{density1}), we have:
\begin{eqnarray}
x_{11} = \frac{1}{Z}\left(a+b\right), \hspace{0.4cm}
x_{12} = \frac{1}{Z}\left( \frac{a}{k_{3}^*}+\frac{b}{k_{4}^*}\right), \hspace{0.4cm}
x_{22} = \frac{1}{Z}\left(\frac{a}{\abs{k_{3}}^{2}}+\frac{b}{\abs{k_{4}}^{2}}\right) \label{x_eqn}
\end{eqnarray}
As a practical matter it is much more convenient to solve for the mean measurement values as a function of $\lambda_{11}$, $\lambda_{12}$ and $\lambda_{22}$. An illustration of this procedure is given in Figure \ref{fig_1}.

\section{IBM Q Test}
To test the performance of our approach, we propose to reconstruct the density matrix in terms of measurements from numerical experiments conducted on IBM quantum computing chips. Several works have already demonstrated reconstructing the density matrix based on measurements from IBM quantum chips. One such work involved characterization of qubit readouts using quantum detector tomography on IBM quantum computers [\cite{chen2019detector}]. IBM Q provides several high-performance simulators to test and optimize any quantum circuit and compare the results with that of real quantum devices. We can compare our reconstruction with the exact density matrix by the quantum circuits that can be derived numerically using the measurements from IBM Q. IBM Q has provided several quantum chips with 5 to 16 qubits available to the public. The IBM device used to carry out calculations for the current work is \textit{IBM Q 5 Yorktown} [\cite{ibm}].

\section{Results and discussion}
The current work includes reconstruction of the density matrix using the known values of a probability (x$_{11}$) and a coherence (x$_{1K}$) and an unknown probability (x$_{KK}$) (for \textit{K}=1,2,$\ldots$,\textit{N}) which is calculated using the available mean measurements. The proposed theory of density matrix reconstruction using maximal entropy formalism is tested and validated for 2-qubit and 3-qubit quantum systems using numerical simulations followed by its implementation on IBM’s Qiskit [\cite{Qiskit}]. 

\subsection{2-Qubit quantum systems}
\subsubsection{Numerical Simulations to test the reconstruction of density matrix}
The 2-qubit quantum system gives rise to SU(4) Lie algebra. In order to reconstruct the density matrix for such a system, given a certain number of known mean measurements, and to verify if such a density matrix is able to reproduce the unknown mean measurement values, let us consider a numerical simulation experiment. For such an experiment, we consider two cases: Case A represents a quantum system for which the expectation values of a probability (x$^A_{11}$) and a coherence (x$^A_{1K}$) are known; Case B represents the same 2-qubit quantum system but with an additional knowledge of one more expectation value (x$^{B}_{KK}$). For both the cases considered, the maximal entropy formalism is employed to calculate the density matrix of the 2-qubit system. The density matrix for the case B serves as the original density matrix and so, if the density matrix reconstructed for case A with fewer known measurements is able to predict the value of the unknown probability (x$^{B}_{KK}$) then it can be shown that the tomography of a quantum system can be carried out using the maximal entropy formalism with the knowledge of fewer known mean measurement values. Let us look at the two cases in more details with specific illustration for \textit{K} = 2 which can then be generalized for \textit{K} = 3, 4 for the 2-qubit quantum systems: \par
\textbf{Case A (x$_{11}^{A}$, x$_{12}^{A}$) and case B (x$_{11}^{B}$, x$_{12}^{B}$, x$_{22}^{B}$) known}\par
\begin{figure}[ht!]
        \centering
        \subcaptionbox{The plot of $x_{22}^{B}$ vs $x_{22}^{A}$ where $x_{22}^{A}$ is predicted using the reconstructed density matrix.}{\includegraphics[width=3.15in]{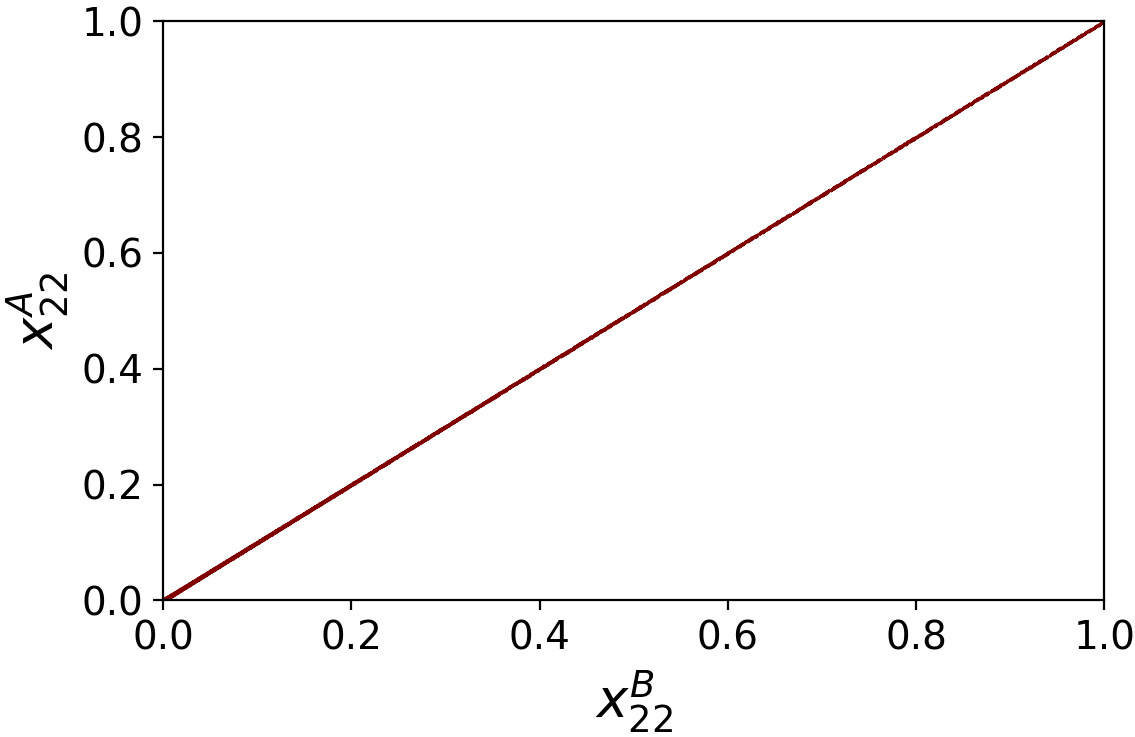} }  \hspace*{\fill}
        \subcaptionbox{Plot of the absolute difference between the predicted and the true mean value $x_{22}$.}{\includegraphics[width=3.15in]{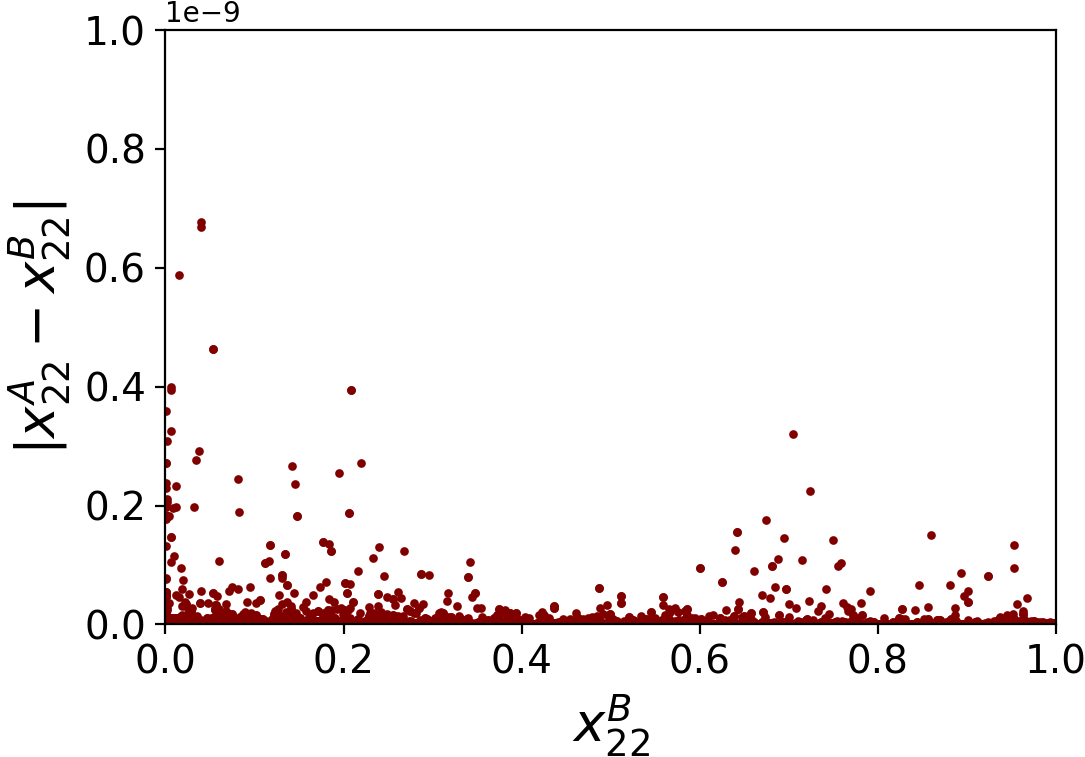}}        
        \caption{ The linear plot in Figure (a) shows that $x_{22}^{A}$ = $x_{22}^{B}$ for all values of $x_{22}^{B}$. Also, absolute differences of the order of 1e-9 in Figure (b) show that the reconstructed density matrix correctly predicts the value of x$_{22}$.}
        \label{fig:my_label1}        
     \end{figure}

To reconstruct the density matrix for both these cases as per Eq. (\ref{density1}) the Lagrange multipliers ($\lambda_{11}$, $\lambda_{12}$, $\lambda_{22}$) need to be determined. For the ease of numerical calculations, the Lagrange multipliers are assumed to be real. For the particular case A, a value for the unknown probability (x$_{22}^A$) is determined using the Appendix Eq. (\ref{x12b}) and then used in the evaluation of Lagrange multipliers. The unknown Lagrange multipliers can easily be calculated from Eq. (\ref{x_eqn}) using the available mean measurements: (x$_{11}$, x$_{12}$, x$_{22}^A$) for case A and (x$_{11}$, x$_{12}$, x$_{22}^B$) for case B. Upon reconstruction of the density matrices, the values of the predicted probability is compared with the true values to substantiate the proposed approach. The linear plot between the true and predicted values of x$_{22}$ in Figure \ref{fig:my_label1}a and their absolute difference plot in Figure \ref{fig:my_label1}b show how close the predicted values (x$_{22}^{A}$) are to the true values (x$_{22}^{B}$) of the mean measurements. Thus, the reconstructed density matrix for a 2-qubit system is successful in accurately predicting x$_{22}$ using just a probability (x$_{11}$) and a coherence (x$_{12}$).

\begin{figure}[ht!]
        \centering
        \subcaptionbox{The plot of $x_{33}^{B}$ vs $x_{33}^{A}$ where $x_{33}^{A}$ is predicted using the reconstructed density matrix.}{\includegraphics[width=3.15in]{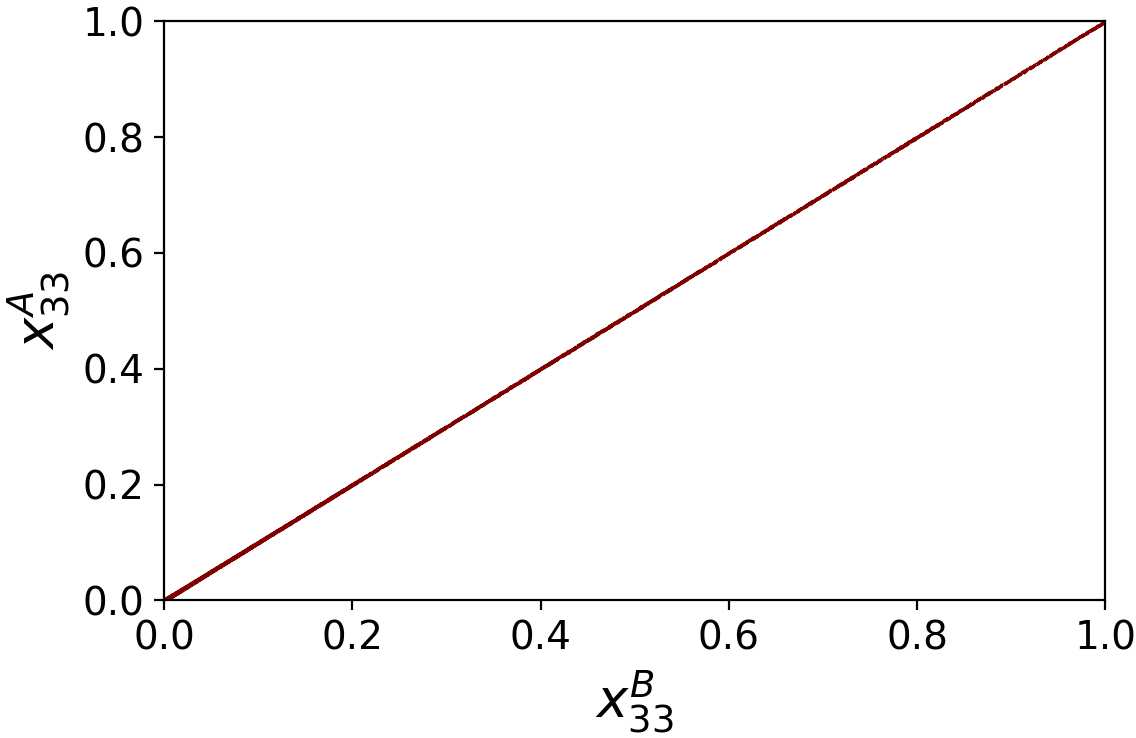} }  \hspace*{\fill}
        \subcaptionbox{Plot of the absolute difference between the predicted $x_{33}^{A}$ and the true mean value $x_{33}^{B}.$}{\includegraphics[width=3.15in]{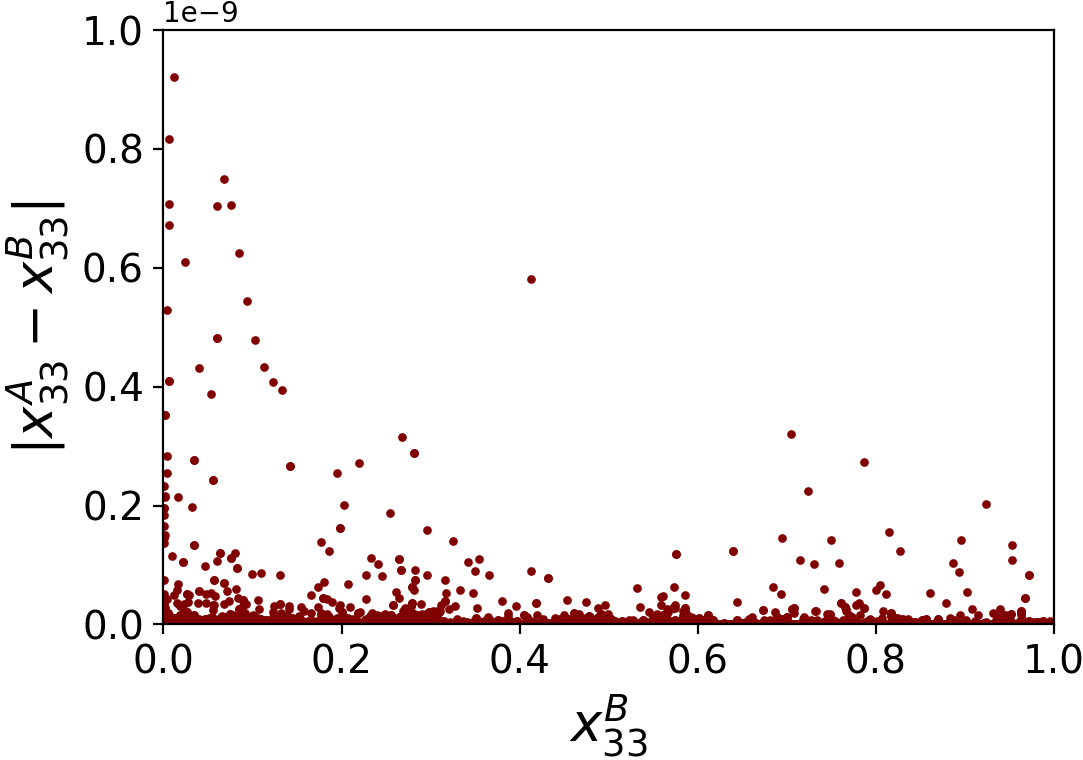}} 
        
        \subcaptionbox{The plot of $x_{44}^{B}$ vs $x_{44}^{A}$ where $x_{44}^{A}$ is predicted using the reconstructed density matrix.}{\includegraphics[width=3.15in]{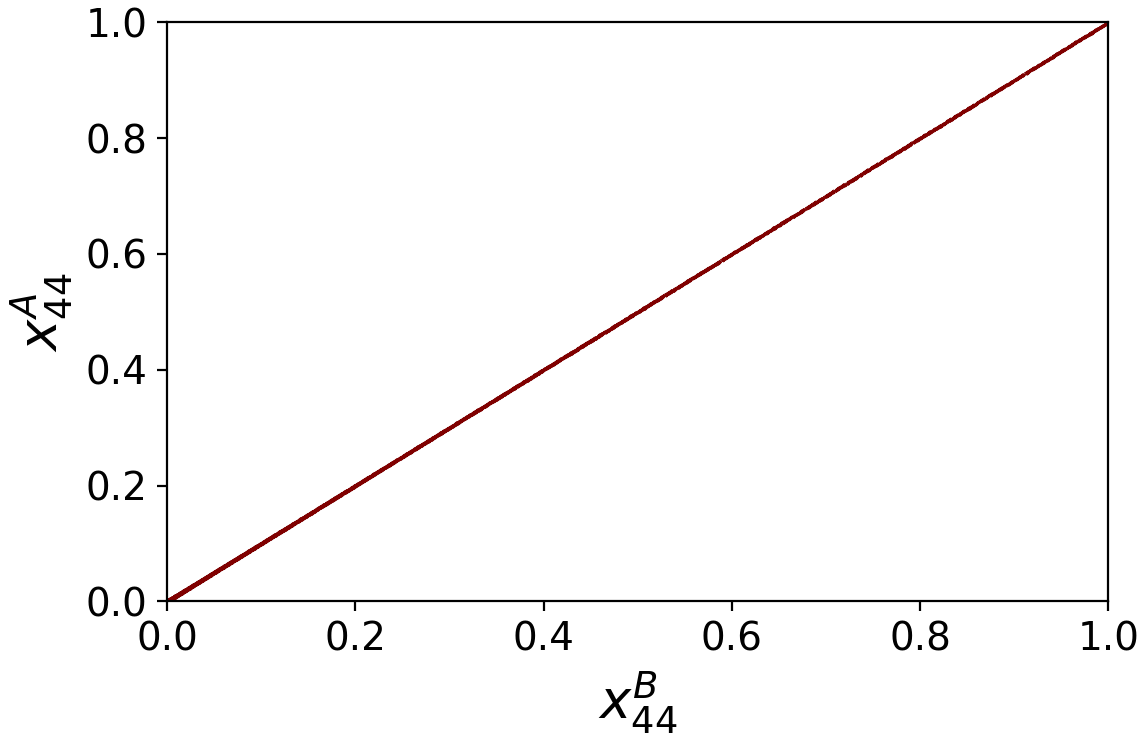}} \hspace*{\fill}
        \subcaptionbox{Plot of the absolute difference between the predicted $x_{44}^{A}$ and the true mean value $x_{44}^{B}.$}{\includegraphics[width=3.15in]{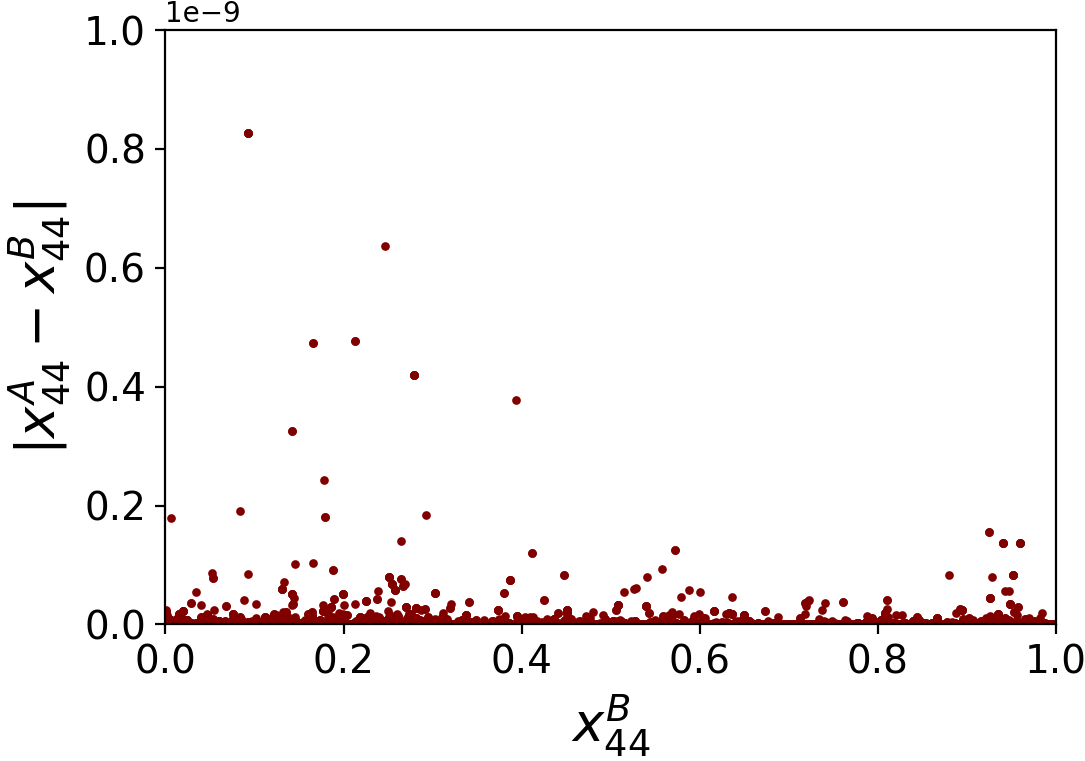}}        
        \caption{ The linear plot in Figure (a) shows that $x_{22}^{A}$ = $x_{22}^{B}$ for all values of $x_{22}^{B}$. Also, absolute differences of the order of 1e-9 in Figure (b) show that the reconstructed density matrix correctly predicts the value of x$_{22}$.}
        \label{fig:my_label12a}        
     \end{figure}

Just like the above case of the prediction of x$_{22}$ for 2-qubit quantum systems, the procedure to reconstruct the density matrix for the prediction of x$_{33}$ or x$_{44}$ is also similar. Here also, to accurately predict x$_{KK}$ we assume that 2 known mean measurements are available: x$_{11}$ and x$_{1K}$. A similar numerical simulation experiment is performed to test the prediction of the unknown observable (x$_{KK}$). As can be seen in Figures \ref{fig:my_label12a}b and \ref{fig:my_label12a}d, the absolute difference between the predicted and the true mean values of x$_{KK}$ is of the order of 1e-9. The numerical simulation results confirm that the maximal entropy formalism can successfully reconstruct the density matrix with fewer than \textit{$N^2$-1} known mean measurements.

        
\subsubsection{Implementation on IBM's Qiskit}
Qiskit is an open-source quantum computing software development framework provided by IBM to run quantum programs on prototype quantum devices in IBM Q. The four different elements of Qiskit: \textit{Terra}, \textit{Aer}, \textit{Ignis} and \textit{Aqua} provide the tools to build and run quantum circuits, simulate device behavior, study and mitigate errors, and solve real-world problems. \\
To test the reconstruction of density matrix we first built various quantum circuits by applying different gate models followed by simulating the circuit with the \textit{Aer} provider in Qiskit. For the Qiskit software stack, the \textit{Aer} provides a high-performance simulator framework with multiple backends to simulate quantum circuits. For the current 2-qubit system, we used three different gate models comprising of Hadamard, Control-X, Control-Z and rotation gates: R$_x$, R$_y$, R$_z$. We vary the rotation angle $\theta$ of the rotation gates to obtain the various quantum circuits to test our reconstruction. Upon construction of the circuit, the \textit{statevector$\_$simulator} backend from \textit{Aer} is used to obtain the final state upon simulation. This serves as a method of obtaining mean measurements without the consideration of errors. From the state vector, one can obtain the mean values of the measurements as follows. Consider the 2-qubit quantum system obtained upon simulation of the circuit as per the aforementioned gate model such that the system can be described by the following state vector:
\begin{eqnarray}
\ket{\psi} = a_{0}\ket{00} + a_{1}\ket{01} + a_{2}\ket{10} + a_{3}\ket{11} \label{state1} \end{eqnarray}
where \textit{a}$_i$’s are the complex amplitudes and $\abs{a_{i}}^{2}$ = $\abs{\braket{i}{\psi}}^{2}$ correspond to the probabilities of the measurements:
\begin{eqnarray}
x_{11} = a_{0}^{*}a_{0} \hspace{0.3cm} ; \hspace{0.3cm} x_{22} = a_{1}^{*}a_{1}\hspace{0.3cm} ;\hspace{0.3cm} x_{33} = a_{2}^{*}a_{2} \hspace{0.3cm};\hspace{0.3cm} x_{44} = a_{3}^{*}a_{3} \label{abcd1}
\end{eqnarray}
Since the complex amplitudes for the 2-qubit states are available from the state vector, we can also calculate the coherences:
\begin{eqnarray}
a_{i}^{*}a_{j} = \bra{\psi}(\ket{i}\bra{j})\ket{\psi} \label{abcd2}
\end{eqnarray}
For example, $x_{12}$ = $a_{0}^{*}a_{1}$; $x_{13}$ = $a_{0}^{*}a_{2}$, etc. \par
\begin{figure}[ht!]
        \centering
        \includegraphics[scale=0.73]{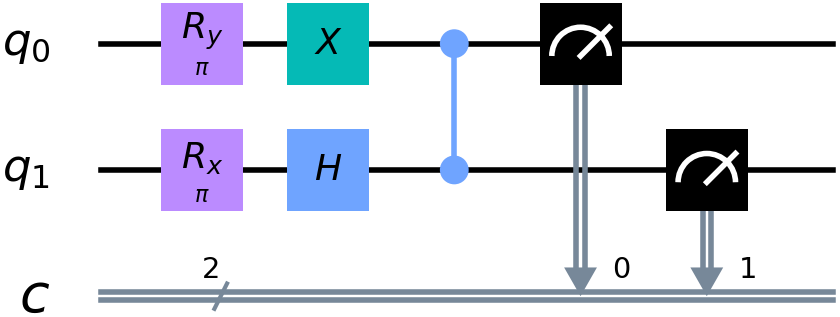} \hspace*{\fill}
        \includegraphics[scale=0.57]{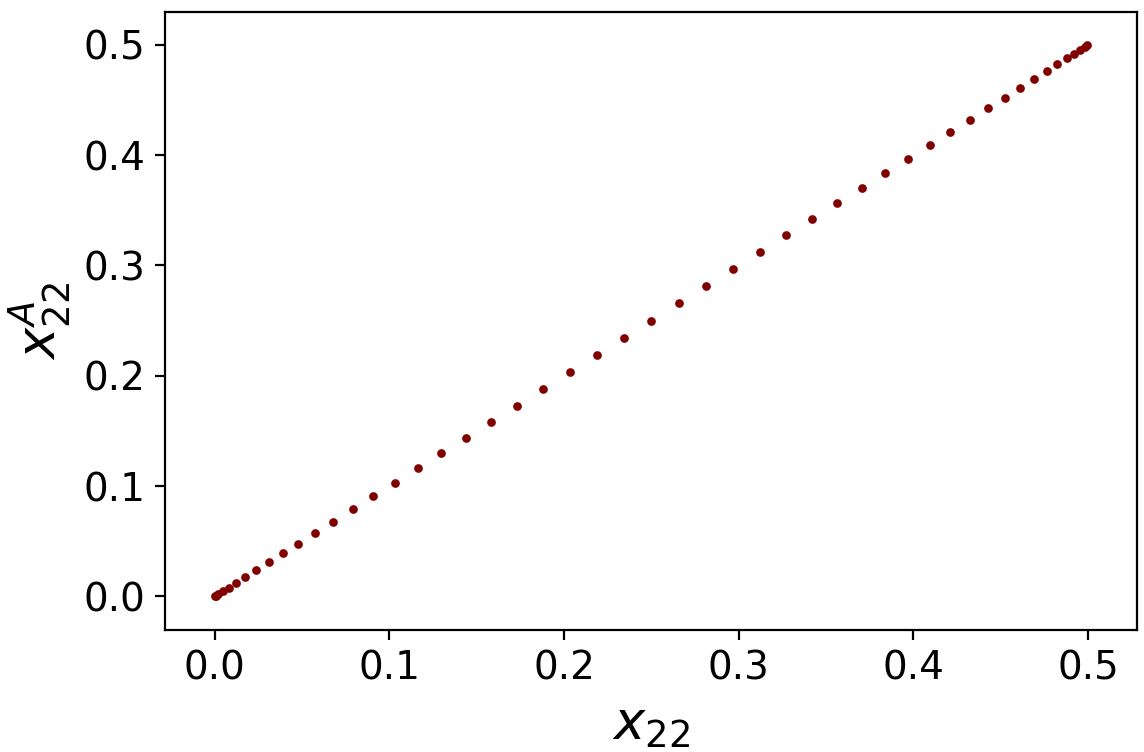} 
        \includegraphics[scale=0.73]{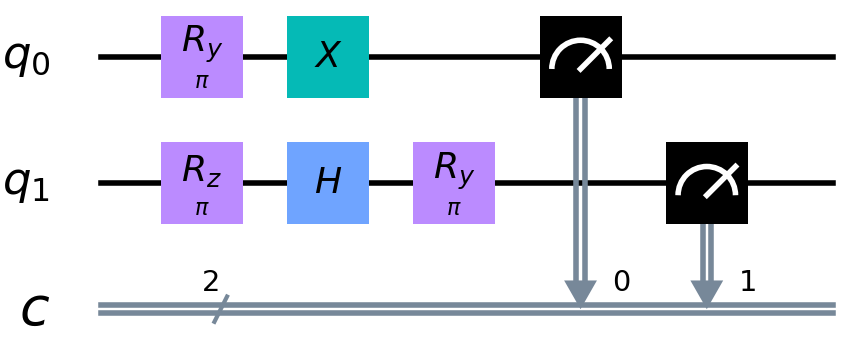} \hspace*{\fill}
        \includegraphics[scale=0.57]{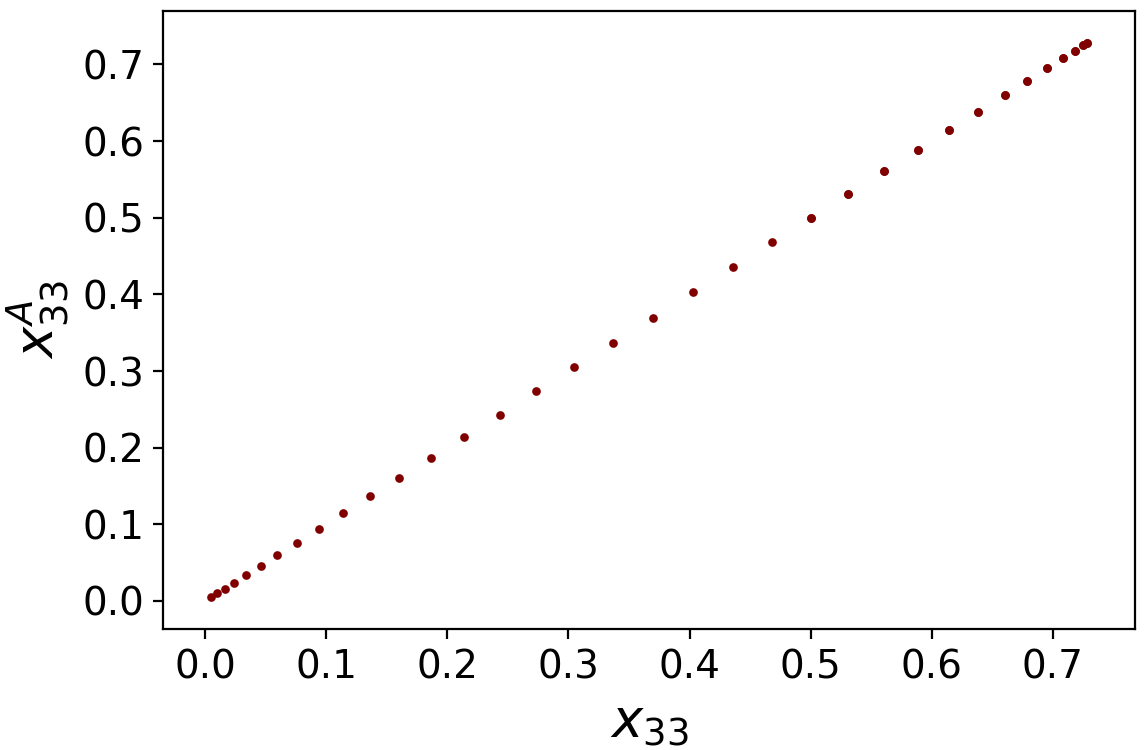} 
        \includegraphics[scale=0.73]{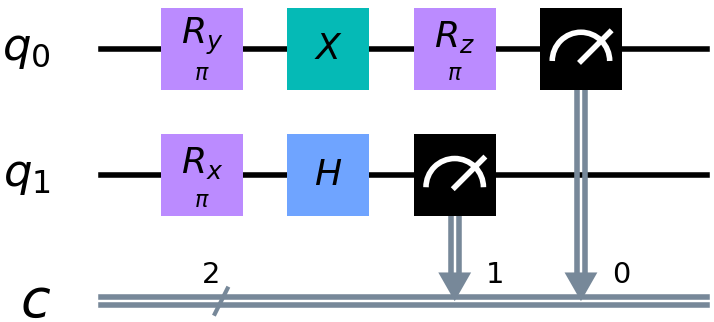} \hspace*{\fill}
        \includegraphics[scale=0.57]{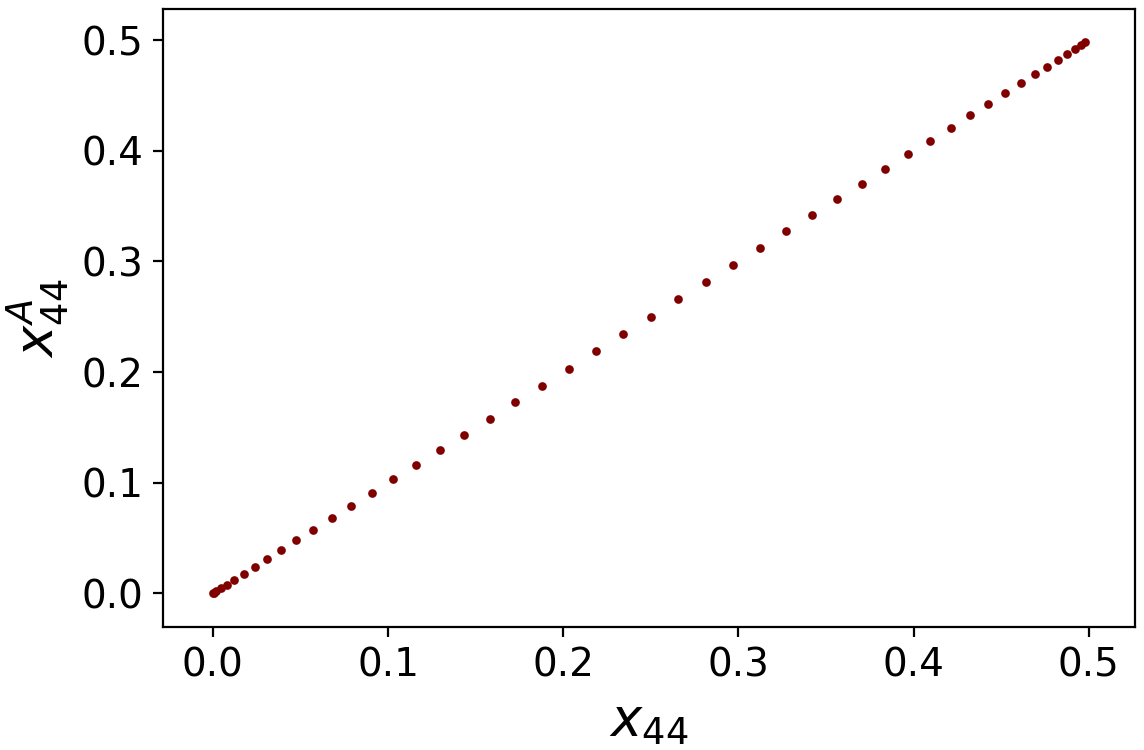} 
        \caption{The different gate models (shown here for one rotation angle $\theta = \pi$) and the corresponding plots between the true (x$_{KK}$) and the predicted (x$_{KK}^A$) mean measurements.}
        \label{fig:my_label5}        
\end{figure}

After obtaining the probabilities and the coherences from the state vector, the mean values of the measurements are calculated using the formalism from maximal entropy theory to compare the results. Figure \ref{fig:my_label5} shows the three gate models along with the corresponding plots between the true and the predicted mean measurements for x$_{KK}$. The linear plots verify that the reconstructed density matrix correctly predicts the value of x$_{KK}$ as obtained from measurements.

\subsection{3-Qubit quantum systems}
The 3-qubit quantum system gives rise to SU(8) Lie algebra. The density matrix reconstruction for such a system using the maximal information entropy formalism is similar to that for the 2-qubit system where the SU(4) group specifies the system. Here also, it is assumed that 2 mean measurements are known (x$_{11}$, x$_{1K}$) and we try to predict x$_{KK}$ using the reconstructed density matrix. The density operator for the 3-qubit system when (x$_{11}$, x$_{12}$, x$_{22}$) are the known mean measurements is:
\begin{eqnarray}
\hat{\rho}&=&\frac{1}{Z}\sum_{i}\exp{\epsilon_{i}}|\phi_{i}><\phi_{i}| \nonumber 
\end{eqnarray}
\begin{eqnarray}
\hat{\rho}&=&\frac{1}{Z}(|8><8|+|7><7|+|6><6|+|5><5|+|4><4| \nonumber \\  
&+&|3><3|+(a+b)|1><1|+(\frac{a}{k_{3}^*}+\frac{b}{k_{4}^*})|1><2| \nonumber\\
&+&(\frac{a}{k_{3}}+\frac{b}{k_{4}})|2><1|+(\frac{a}{\abs{k_{3}}^{2}}+\frac{b}{\abs{k_{4}}^{2}})|2><2|) \label{abc}] 
\end{eqnarray}
where Z=$\sum_{i}\exp{\epsilon_{i}}$, k$_3$ = -($\frac{\epsilon_3}{\lambda_{12}^{*}}$+$\frac{\lambda_{22}}{\lambda_{12}^{*}}$), k$_4$ = -($\frac{\epsilon_4}{\lambda_{12}^{*}}$+$\frac{\lambda_{22}}{\lambda_{12}^{*}}$), a=$\frac{\abs{k_{3}}^{2}}{\sqrt{(k_3^2+1)({k_3^*}^2+1)}}\exp{\epsilon_{3}}$, and \par b=$\frac{\abs{k_{4}}^{2}}{\sqrt{(k_4^2+1)({k_4^*}^2+1)}}\exp{\epsilon_{4}}$. \par
The method adopted to validate the theory for the SU(8) case is same as that for the SU(4) case. Upon simulation of the quantum circuit using the \textit{statevector$\_$simulator} backend, we obtain the state vector corresponding to the final state:
\begin{eqnarray}
\ket{\psi} = a_{0}\ket{000} + a_{1}\ket{001} + a_{2}\ket{010} + a_{3}\ket{011} + a_{4}\ket{100} + a_{5}\ket{101} + a_{6}\ket{110} + a_{7}\ket{111}
\end{eqnarray}
The mean values of the measurements: the probabilities and the coherences are calculated using Eq. (\ref{abcd1}) and Eq. (\ref{abcd2}) respectively which are used to reconstruct the density matrix for various quantum circuits and predict the unknown probability x$_{KK}$ as discussed in the following section for one example gate model.

\subsubsection{Testing on some complicated circuits}
\begin{figure}[ht!]
        \centering
        \includegraphics[scale=0.5]{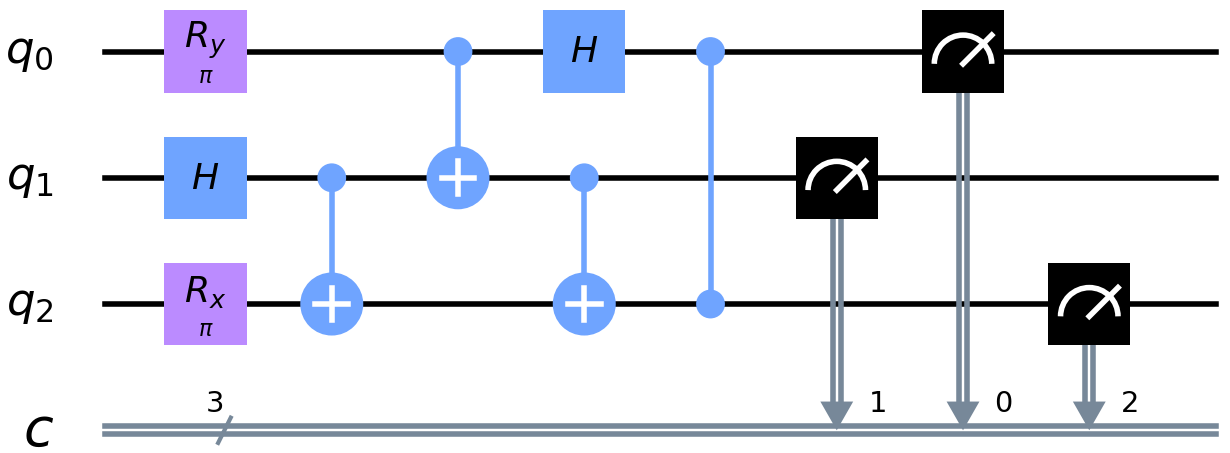}\hspace*{\fill}
        \includegraphics[width=3.15in]{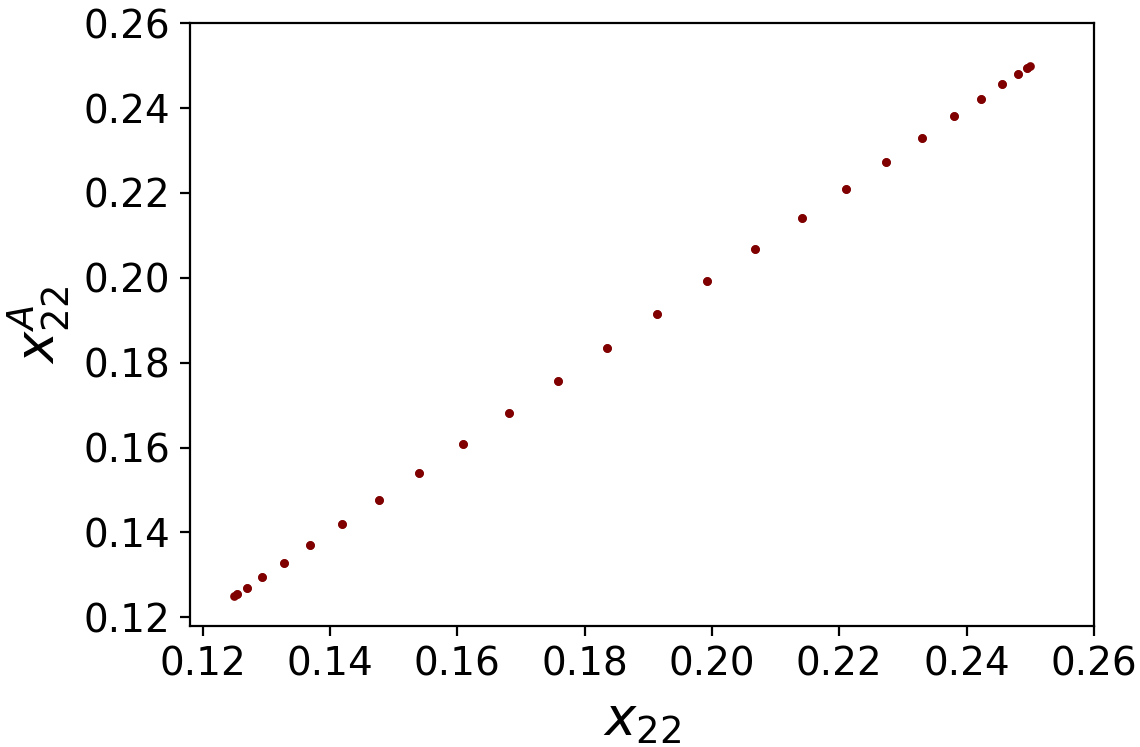}
        \includegraphics[width=3.15in]{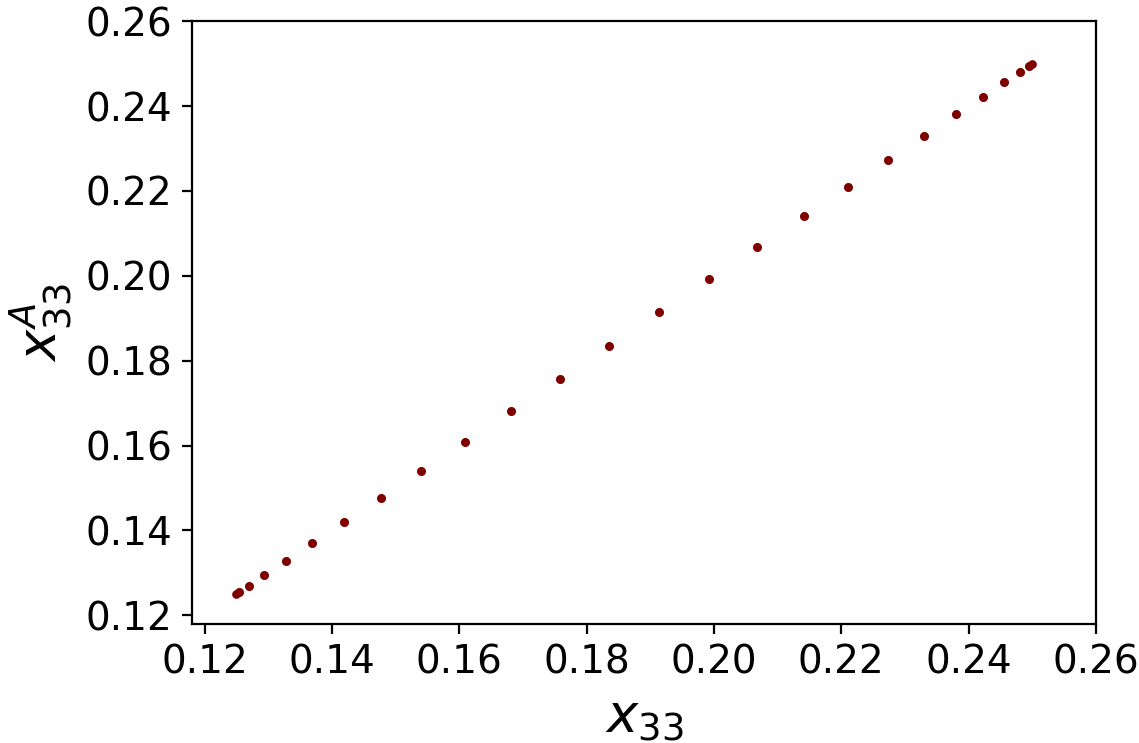}\hspace*{\fill}
        \includegraphics[width=3.15in]{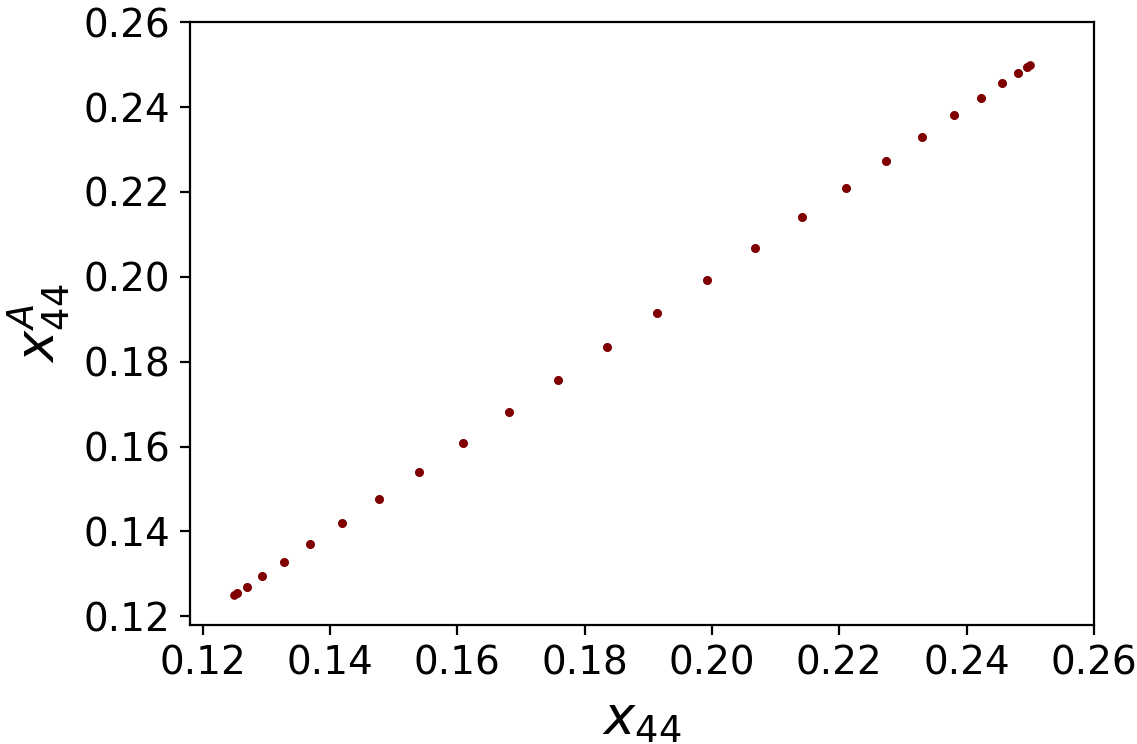}
        \caption{A sample circuit diagram with variable $\theta$ in R$_{x}$ and R$_{y}$ gates for 3-qubit SU(8) case and the corresponding plots of the true (x$_{KK}$) and the predicted (x$_{KK}^A$) mean values for x$_{KK}$. The linear plots show that the maximal entropy formalism-based reconstructed density matrix accurately predicts x$_{KK}$.}
        \label{fig:my_label10}        
\end{figure}

Figure \ref{fig:my_label10} shows an example quantum circuit with a gate model comprising of various quantum gates such as Hadamard, Control-X, Control-Z, R$_x$, and R$_y$ gates. The rotation angle $\theta$ in the R$_x$ and R$_y$ gates on qubit 0 and qubit 2 is varied to obtain the different quantum circuits. Upon construction of the circuits, the \textit{statevector$\_$simulation} is performed to obtain the mean values of measurements and the corresponding values for x$_{KK}$ are predicted using the reconstructed density matrix. It can be seen in Figure \ref{fig:my_label10} that the results for x$_{KK}$ obtained from the simulations match with those predicted by the reconstructed density matrix.

\section{Error Analysis}
Three techniques to generate inputs for obtaining the mean values of measurements are considered to verify the reconstructed density matrix for the prediction of x$_{KK}$ for 2-qubit and 3-qubit quantum systems. For the first technique, the \textit{statevector$\_$simulator} backend of the \textit{Aer} element in qiskit is employed as discussed in Eq. (\ref{state1}). The advantage of this approach is that the coherence can be calculated directly from the complex amplitudes of the basis states. Although, this would not help in the estimation of error as from any experiment on a real quantum device the exact state cannot be obtained. However, we could verify that the maximal entropy formalism-based density matrix reconstruction accurately predicts the value of the unknown probability by comparing the expectation values from the reconstructed density matrix with the expectation values from the \textit{statevector$\_$simulator} backend. \par
The second technique of generating inputs employs the \textit{qasm$\_$simulator} backend in the \textit{Aer} provider in Qiskit wherein only the probabilities can be obtained directly and to obtain the coherences, the corresponding operator is decomposed into tensor products of Pauli matrices as discussed in the Appendices. The calculation of coherences using the decomposition method results in statistical errors which is inversely proportional to the square root of the number of times a circuit is run on the simulator. These statistical errors are reflected in the results of the simulation of circuit using the \textit{qasm$\_$simulator} backend. \par
Finally, the density matrix is reconstructed using the obtained mean measurement values from the IBM machine for the considered quantum circuit using the 5-qubit quantum chip with \textit{ibmqx2} backend [\cite{ibm}]. From the IBM machine experiment also, we can only obtain the probabilities and therefore, again the coherences are obtained using the decomposition method described in the Appendices. Apart from this, we also used qiskit’s \textit{ignis.mitigation.measurement} module in order to mitigate the measurement errors. This is carried out by constructing a calibration matrix for the measurements and thereby mitigating the measurement errors. \par
\begin{figure}[ht!]
        \centering
        \includegraphics[scale=0.6]{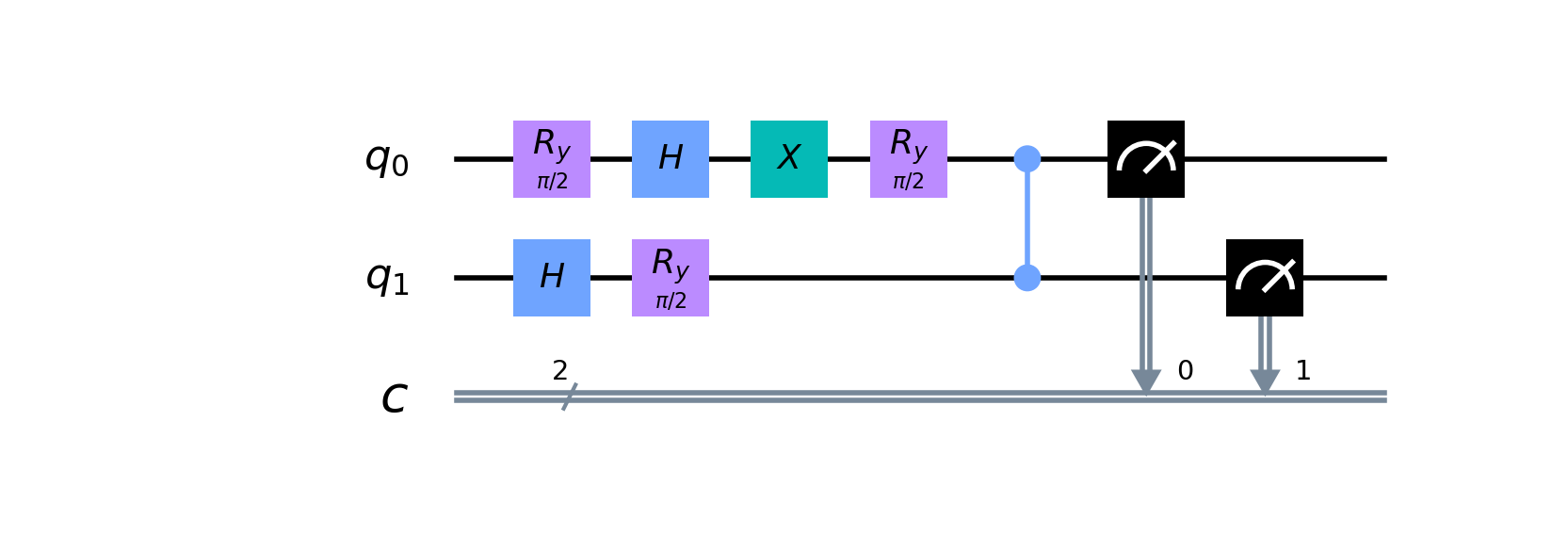}
        \includegraphics[width=3.2in]{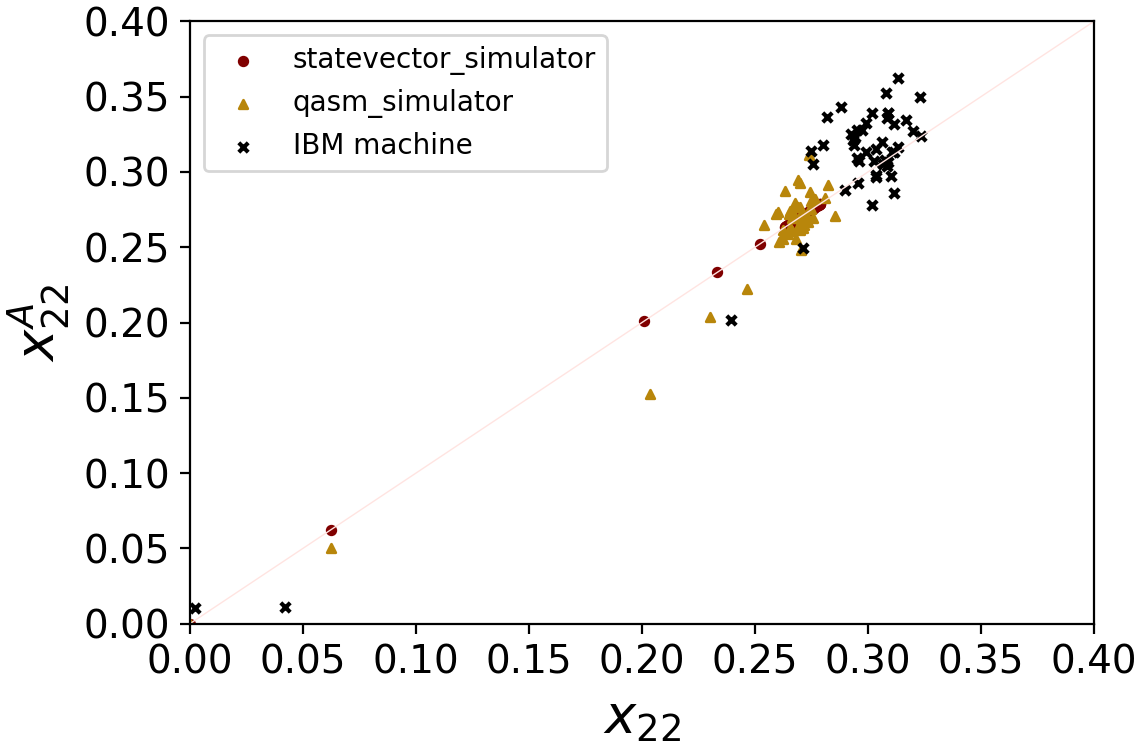} \hspace*{\fill}
        \includegraphics[width=3.2in]{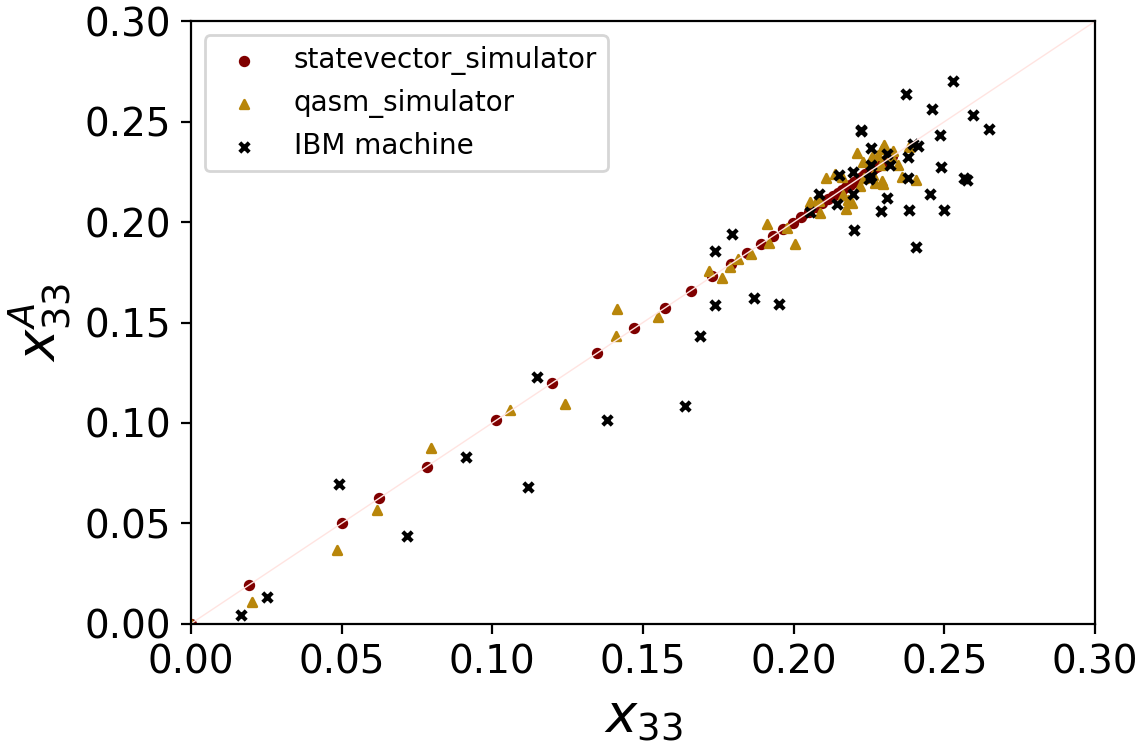}
        \caption{Plots of the true (x$_{22}$) vs the predicted (x$_{22}^A$) mean measurement values of x$_{22}$ and x$_{33}$ for the 2-qubit SU(4) case for the shown gate model using the three backends: \textit{statevector$\_$simulator}, \textit{qasm$\_$simulator} and \textit{ibmqx2}. The reconstructed density matrix accurately predicts x$_{22}$ and x$_{33}$ in the case of \textit{statevector$\_$simulator} whereas slight deviations are observed in the case of \textit{qasm$\_$simulator} method and IBM machine experiment. }
        \label{fig:my_label14}        
\end{figure}

As can be seen in Figure \ref{fig:my_label14}, for the sample circuit considered the maximal entropy formalism-based density matrix reconstruction accurately predicts the mean measurement values of x$_{22}$ and x$_{33}$ in the case of \textit{statevector$\_$simulator} for 2-qubit SU(4) scenario. In the case of \textit{qasm$\_$simulator}, the proposed formalism approximately predicts the mean measurement value. This is what we expect upon measurement of the considered 2-qubit quantum states, up to statistical fluctuations. Furthermore, the IBM machine experiment’s results also show that the predicted mean measurement values of x$_{22}$ and x$_{33}$ for the various circuits are very close to the true values which correspond to the mean values of x$_{22}$ and x$_{33}$ obtained upon execution of the circuit on the IBM machine.

\section{Concluding Remarks}
We discussed generating a reliable and practical inference for the quantum state of a system. We paid special reference to circumstances where the information on the system is not only the probabilities of possible outcomes but also coherences between different outcomes. The need for such an inference arises in a variety of contexts. We have emphasized applications to validate the operation of noisy intermediate-scale quantum devices. There are other circumstances where the very measurement of probabilities and coherences, e.g., [\cite{fresch},\cite{Gattuso2020},\cite{fresch2}], is of primary interest. There too, because of noise, not all the variables of interest can be measured. As we show, partial measurements can still be very useful because the inferred state can be used to compute the mean values of the missing variables. \par
Although our current approach is most suitable for noisy intermediate-scale quantum devices, the simplicity and predicting power  of this alternative approach to quantum tomography, based on the maximal information entropy, invites generalization to other SU(N) systems.  Research is underway to generalize this approach to larger density matrices and analyze the scaling complexity of this approach to the general case of SU(N).

\setcounter{secnumdepth}{0}
\section{Acknowledgment}
We acknowledge the partial financial support by the U.S. Department of Energy (Office of Basic Energy Sciences) under Award No. DE-SC0019215. We also acknowledge use of the IBM Q for this work. The views expressed here are those of the authors and do not reflect the official policy or position of IBM or the IBM Q team.

\clearpage

\bibliography{achemso-demo}

\clearpage
\let\appendixpagenameorig\appendixpagename
\renewcommand{\appendixpagename}{\LARGE\appendixpagenameorig}
\setcounter{secnumdepth}{1}
\begin{appendices}
\renewcommand{\theequation}{A.\arabic{equation}}
\setcounter{equation}{0}
\section{Detailed reconstruction of the density matrix for SU(4) group}
Consider a quantum system which gives rise to SU(4) Lie group algebra. To reconstruct the density matrix for such a system from a finite set of observables let us assume that we know x$_{11}$, x$_{12}$ and x$_{22}$. For the complete description of the system, we require a basis of 16 operators:
\begin{eqnarray}
    &\hspace*{0.001cm}&\{\ket{1}\bra{1},\ket{2}\bra{2},\ket{3}\bra{3},\ket{4}\bra{4},(\ket{1}\bra{2},\ket{2}\bra{1}),(\ket{1}\bra{3},\ket{3}\bra{1}),(\ket{1}\bra{4},\ket{4}\bra{1}), \nonumber \\
    &\hspace*{0.001cm}&(\ket{2}\bra{3},\ket{3}\bra{2}), (\ket{2}\bra{4},\ket{4}\bra{2}), (\ket{3}\bra{4},\ket{4}\bra{3}) \} \label{basis}    
\end{eqnarray}
Based on the maximal entropy formalism we can write the density operator in terms of the operators corresponding to the available observables as:
\begin{eqnarray}
\hat{\rho} = \frac{1}{Z(\lambda_{11},\lambda_{12},\lambda_{22})}\exp\{-\lambda_{11}\ket{1}\bra{1}-\lambda_{12}\ket{1}\bra{2} -\lambda_{12}^{*}\ket{2}\bra{1}-\lambda_{22}\ket{2}\bra{2}\} 
\end{eqnarray}
In this case the operators do not commute. To compute the density matrix we first diagonalize the matrix \textbf{A} which we get upon summing the terms in the exponent within the SU(4) representation:
\[
A = 
\begin{bmatrix}
 -\lambda_{11} & -\lambda_{12} & 0 & 0  \\
  -\lambda_{12}^{*} & -\lambda_{22} & 0 & 0 \\
  0 & 0 & 0 & 0 \\
  0 & 0 & 0 & 0 
 \end{bmatrix} =\sum_{i=1}^{4} \epsilon_{i}\ket{\phi_{i}}\bra{\phi_{i}}
\]
with $\{\epsilon_i,\ket{\phi_i}\}$ being eigenvalues and corresponding eigenvectors of the matrix \textbf{A}:
\begin{eqnarray}
\epsilon_1 = 0 &;& \bra{\phi_1} = (0\hspace{0.2cm}0\hspace{0.2cm}0\hspace{0.2cm} 1) \nonumber \\
\epsilon_2 = 0 &;& \bra{\phi_2} = (0\hspace{0.2cm}0\hspace{0.2cm}1\hspace{0.2cm} 0) \nonumber \\
\epsilon_3 = -\frac{1}{2}(\lambda_{11}+\lambda_{22}+\sqrt{4\lambda_{12}\lambda_{12}^{*}+\lambda_{11}^2+\lambda_{22}^2-2\lambda_{11}\lambda_{22}} ) &;& \bra{\phi_3} = (k_3\hspace{0.2cm}1\hspace{0.2cm}0\hspace{0.2cm} 0) \nonumber \\
\epsilon_4 = -\frac{1}{2}(\lambda_{11}+\lambda_{22}-\sqrt{4\lambda_{12}\lambda_{12}^{*}+\lambda_{11}^2+\lambda_{22}^2-2\lambda_{11}\lambda_{22}} ) &;& \bra{\phi_4} = (k_4\hspace{0.2cm}1\hspace{0.2cm}0\hspace{0.2cm} 0) \nonumber
\end{eqnarray}
where k$_3$ = -($\frac{\epsilon_3}{\lambda_{12}^{*}}$+$\frac{\lambda_{22}}{\lambda_{12}^{*}}$), k$_4$ = -($\frac{\epsilon_4}{\lambda_{12}^{*}}$+$\frac{\lambda_{22}}{\lambda_{12}^{*}}$) \newline
Now we can compute the density operator in the basis of the eigenvectors of \textbf{A}:
\begin{eqnarray}
\hat{\rho} &=& \frac{1}{Z}\big( \exp{\epsilon_1}\ket{\phi_1}\bra{\phi_1}+\exp{\epsilon_2}\ket{\phi_2}\bra{\phi_2}+\exp{\epsilon_3}\ket{\phi_3}\bra{\phi_3}+\exp{\epsilon_4}\ket{\phi_4}\bra{\phi_4} \big) \nonumber \\
Z &=& tr\big( \exp{\textbf{A}} \big) = \sum_{i=1}^4 \exp{\epsilon_i} \nonumber
\end{eqnarray}
As far as we computed the exponent of the operators, now we can expand the projection operators $\ket{\phi_i}\bra{\phi_i}$ in our initial basis (\ref{basis}):
\begin{eqnarray}
|\phi_{1}><\phi_{1}| &=& |4><4| \nonumber \\
|\phi_{2}><\phi_{2}| &=& |3><3| \nonumber \\
|\phi_{3}><\phi_{3}| &=& \frac{1}{\sqrt{(k_3^2+1)({k_3^*}^2+1)}}(\abs{k_{3}}^{2}|1><1|+k_{3}|1><2|+k_{3}^*|2><1|+|2><2|) \nonumber \\
|\phi_{4}><\phi_{4}| &=& \frac{1}{\sqrt{(k_4^2+1)({k_4^*}^2+1)}}(\abs{k_{4}}^{2}|1><1|+k_{4}|1><2|+k_{4}^*|2><1|+|2><2|) \nonumber
\end{eqnarray}
Giving the final form of the density operator:
\begin{eqnarray}
\hat{\rho}&=&\frac{1}{Z}\sum_{i}\exp{\epsilon_{i}}|\phi_{i}><\phi_{i}| \nonumber \\
&=&\frac{1}{Z}(|4><4|+|3><3|+(a+b)|1><1|+(\frac{a}{k_{3}^*}+\frac{b}{k_{4}^*})|1><2| \nonumber\\
&+&(\frac{a}{k_{3}}+\frac{b}{k_{4}})|2><1|+(\frac{a}{\abs{k_{3}}^{2}}+\frac{b}{\abs{k_{4}}^{2}})|2><2|) \label{density}
\end{eqnarray}
where Z=$\sum_{i}\exp{\epsilon_{i}}$, k$_3$ = -($\frac{\epsilon_3}{\lambda_{12}^{*}}$+$\frac{\lambda_{22}}{\lambda_{12}^{*}}$), k$_4$ = -($\frac{\epsilon_4}{\lambda_{12}^{*}}$+$\frac{\lambda_{22}}{\lambda_{12}^{*}}$), a=$\frac{\abs{k_{3}}^{2}}{\sqrt{(k_3^2+1)({k_3^*}^2+1)}}\exp{\epsilon_{3}}$, and \newline b=$\frac{\abs{k_{4}}^{2}}{\sqrt{(k_4^2+1)({k_4^*}^2+1)}}\exp{\epsilon_{4}}$.  To determine the values of the Lagrange multipliers $\lambda_{ij}$ we use the information about the measured mean values of the operators. We define x$_{11}$ = $\langle\ket{1}\bra{1}\rangle$, x$_{12}$ = $\langle\ket{1}\bra{2}\rangle$, and x$_{22}$ = $\langle\ket{2}\bra{2}\rangle$.
Having the density operator defined by (\ref{density}) we obtain:
\begin{eqnarray}
x_{11} = \frac{1}{Z}\big(a+b\big), \hspace{0.5cm} x_{12} = \frac{1}{Z}\big(\frac{a}{k_{3}^*}+\frac{b}{k_{4}^*}\big), \hspace{0.5cm} x_{22} = \frac{1}{Z}\big(\frac{a}{\abs{k_{3}}^{2}}+\frac{b}{\abs{k_{4}}^{2}}\big) \label{xvalues}
\end{eqnarray}
The unknown Lagrange multipliers can be determined by solving the last three equations. These Lagrange multipliers are then used to reconstruct the density matrix based on maximal information entropy formalism.

\section{Analytical expression to calculate probabilities x$_{KK}$ }
Consider a quantum system represented by a state vector $\ket{\psi} = h_1\ket{1}+h_2\ket{2}+h_3\ket{3}+\ldots$ such that $h_1$ = a+\textit{i}b, $h_2$ = c+\textit{i}d, etc. Let the expectation values of the observables be defined in the following way:
\begin{eqnarray}
x_{11} &=& \langle \ket{1}\bra{1}\rangle = a^2 + b^2 \nonumber \\
x_{22} &=& \langle \ket{2}\bra{2}\rangle = c^2 + d^2 \nonumber \\
x_{12}&=& \langle \ket{1}\bra{2}\rangle = (ac+bd)+i(ad-bc) \nonumber \\
x_{21} &=& \langle \ket{2}\bra{1}\rangle = (ac+bd)+i(bc-ad) \label{x21A}
\end{eqnarray}
This leads to an interesting relationship between x$_{11}$, x$_{22}$ and x$_{12}$ which can be used to determine the unknown observable. We have:
\begin{eqnarray}
x_{11} x_{22} &=& a^2c^2 + a^2d^2 + b^2c^2 + b^2d^2 \nonumber \\
\abs{x_{12}}^2 = \abs{x_{21}}^2 &=& a^2c^2 + a^2d^2 + b^2c^2 + b^2d^2 \nonumber \\
\abs{x_{12}}^2 &=& x_{11} x_{22} \label{semi}
\end{eqnarray}
Within the context of this work, the following relation is used to calculate a value for the unknown probability (x$_{22}^A$):
\begin{eqnarray}
x_{22}^A = \frac{\abs{x_{12}}^2}{x_{11}} \label{x12b}
\end{eqnarray}
Therefore, the unknown Lagrange multipliers are determined using the available mean measurements and the calculated probability values using Eq. (\ref{xvalues}) and thereby, the density matrix is reconstructed based on maximum entropy formalism. \newline
Following the above procedure and considering the expectation values of x$_{11}$ and x$_{1K}$ we can obtain a similar expression for the calculation of x$_{KK}^A$ when x$_{11}$ and x$_{1K}$ are the known mean measurements:
\begin{eqnarray}
x_{KK}^A = \frac{\abs{x_{1K}}^2}{x_{11}} \label{x13b}
\end{eqnarray}

\section{A possible approach for coherence measurements on IBM quantum computer}
Considering a 2-qubit (SU(4)) case, it is easy to get the value of $x_{11} = \langle|1\rangle\langle1|\rangle$ etc., by the measurement results from IBM quantum computer because $x_{11}$ corresponds to the probability of getting final results as $|1\rangle$. However, it is not easy to get the results of the coherence on a quantum computer, for example $x_{12} = \langle|1\rangle\langle2|\rangle$, because the operator of coherence is not available at measurements. However, this measurement can be achieved by decomposing the operator into sum of tensor products of Pauli matrices and evaluating each tensor product on the quantum computer. Consider a simple example of evaluating $x_{12} = \langle|1\rangle\langle2|\rangle$. Qiskit's tensor product order is followed to illustrate the considered example.

\section*{Decomposition}
We can rewrite $|1\rangle\langle2|$ as a matrix:
\[
|1\rangle\langle2|=
\begin{bmatrix}
0 & 1 & 0 & 0\\
0 & 0 & 0 &0\\
0 & 0 & 0 & 0\\
0 & 0 & 0 & 0\\
\end{bmatrix}
\]
Considering four 2$\times$2 matrices:
\[
A = \frac{1}{2} (Z + I) = 
\begin{bmatrix}
1 & 0\\
0 & 0 \\
\end{bmatrix}
\]
\[
B = \frac{1}{2} (Z - I) = 
\begin{bmatrix}
0 & 0\\
0 & 1 \\
\end{bmatrix}
\]
\[
C = \frac{1}{2} (X + iY) = 
\begin{bmatrix}
0 & 1\\
0 & 0 \\
\end{bmatrix}
\]
\[
D = \frac{1}{2} (X - iY) = 
\begin{bmatrix}
0 & 0\\
1 & 0 \\
\end{bmatrix}
\]
where $X$, $Y$, $Z$ are Pauli matrices and $I$ is the 2$\times$2 identity matrix. We can get:
\[
\begin{bmatrix}
0 & 1 & 0 & 0\\
0 & 0 & 0 &0\\
0 & 0 & 0 & 0\\
0 & 0 & 0 & 0\\
\end{bmatrix} = A \otimes C \hspace*{1cm}
\]
or
\begin{equation}
    \begin{aligned}
        |1\rangle\langle2| &=&  A \otimes C = \frac{1}{2}Z_2X_1 
    \end{aligned}
\end{equation}

Thus we have successfully decomposed the coherence operator into the  tensor products of the Pauli matrices. Generally speaking, we can use the technique to decompose any coherence operator into the tensor products of the Pauli matrices. The evaluation of each tensor product of Pauli matrices can be achieved by using the rotation gates. More details about this technique can be found in the work [\cite{kandala2017hardware},\cite{bian2019quantum}].

\end{appendices}

\end{document}